\documentstyle[prb,aps,twocolumn,epsf]{revtex}

\def\prb{Phys.\ Rev.\ {\bf B}}
\def\prl{Phys.\ Rev.\ Lett.\/}

\def\be{\begin{equation}}
\def\ee{\end{equation}}
\def\ba{\begin{eqnarray}}
\def\ea{\end{eqnarray}}

\def\C60{A$_x$C$_{60}$}

\def\ie{ {\it i.\ e.\/} }

\def\etal{ {\it et.\ al.\/} }

\begin{document}

\twocolumn[\hsize\textwidth\columnwidth\hsize\csname@twocolumnfalse\endcsname

\title
{Liquid Crystal Phases of Quantum Hall Systems}

\author{Eduardo Fradkin$^{a}$ and Steven A.~Kivelson$^{b}$}
\address{
Department of Physics, University of Illinois$^{a}$, Urbana, IL 
61801-3080
and Department of Physics, U.\ C.\ L.\ A.\ $^{b}$ , Los Angeles, CA  
90095}
\date{\today}
\maketitle
\begin{abstract}
Mean-field calculations for the two dimensional electron gas (2DEG) 
in a 
large magnetic field with a partially filled Landau level
with index $N\geq 2$ consistently yield ``stripe-ordered'' 
charge-density wave ground-states, for much the same reason that 
frustrated phase separation leads to stripe ordered states in doped 
Mott insulators.  We have studied the effects of quantum and 
thermal
fluctuations about such a state and show that they can lead to a set 
of 
electronic liquid crystalline states, particularly a stripe-nematic 
phase which is stable at $T>0$. 
Recent measurements of the longitudinal resistivity of a set of
quantum Hall devices have revealed that these
systems spontaneously develop, at low temepratures,
a very large anisotropy.  We interpret these experiments as evidence 
for a stripe nematic phase, and propose a general phase diagram for 
this system.
\end{abstract}

\

\

]

\narrowtext

There are many condensed matter systems in which the charge degrees 
of 
freedom form regular spacial patterns commonly known as ``stripes.'' 
These structures are
typically the result of the competition between short range 
attractive forces,
which give rise to a condensed (usually insulating) phase, and the 
largely unscreened, 
long range Coulomb interactions.  Specifically,
in the condensed phase, and in the absence of long range repulsive 
forces between
like charges, the charge degrees of freedom have a tendency to form 
clumps, that is to
phase separate. This tendency to phase separation is frustrated by 
the long range 
repulsive Coulomb interactions and the result is the spontaneous 
organization of the
charge degrees of freedom in low dimensional structures\cite{topo}. In
a large class of quasi two-dimensional strongly correlated materials, 
\ie doped Mott insulators such as the copper
oxides, the nickelates, and the manganates, stripe phases have 
recently been 
observed experimentally.\cite{tranq,yamada} 

However,  these structures should not be peculiar to doped Mott 
insulators,
but should also arise in other electronic systems in which the same
sort of competition is present.  It has been realized for some time 
that there is a strong tendency for the two dimensional electron gas 
(2DEG) in a high magnetic field to condense into incompressible 
quantum Hall liquid states with quantized ``filling factor''
$\nu$.  Thus, in the presence of long-range Coulomb interactions,
at electron densities intermediate between two 
quantized values it is natural to expect the system to form an 
inhomogeneous state with a periodic array of
stripes of the two incompressible liquids. Indeed,
some time ago Kulakov, Fogler and Shklovskii\cite{kfs} and Moessner 
and
Chalker\cite{chalker} showed that a two-dimensional electron gas in a
perpendicular magnetic field can have a stripe or 
charge
density wave (CDW) ground state in which the charge density in a
partially-filled high
Landau level exhibits periodic oscillations along one spacial direction. 
According to these
calculations, which are based on a Hartree-Fock approach, the 
electrons in a 
partially occupied $N^{\rm th}$ Landau 
level form stripes
in which the Landau level is alternately full or empty. The stripe 
pattern 
is 
a periodic modulation of the local Hall conductance between the 
quantized values $2N
{\frac{e^2}{h}}$ and $(2N+1){\frac{e^2}{h}}$. 

\begin{figure}
\begin{center}
\leavevmode
\epsfxsize=3in
\epsfbox{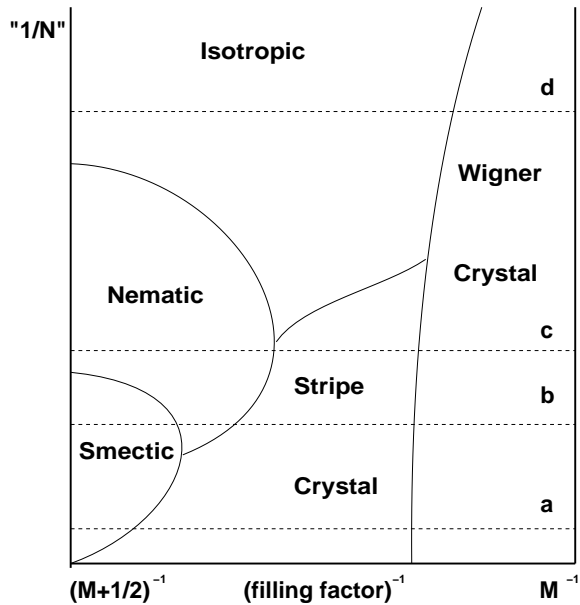}
\end{center}
\caption
{Qualitative $T=0$ phase diagram for a clean 2DEG. The vertical axis
measures the strength of 
the quantum fluctuations (which is roughly inversely proportional to 
the Landau level index, $N$) and the horizontal axis is the inverse 
filling
factor (over a range in which the partial filling of the highest
spin-polarized Landau level varies between $M$ and $M+1/2$, where
$M=2N$ or $M=2N+1$.  Lines a, b, c and d 
are three realizations for increasing the strength of the quantum 
fluctuations.  The Smectic, 
and Nematic are compressible while the Isotropic phase may either be 
compressible or 
incompressible. 
The Stripe Crystal and the Wigner Crystal phases are insulating 
and exhibit 
plateau behavior with the same quantized Hall conductance, 
$\sigma_{xy}=
e^{2}M/h$. }
\label{fig:fig2}
\end{figure}

In this paper we present a theoretical description 
of the 2DEG 
in a moderately large magnetic field, 
and show that this system actually behaves like a set of
dynamical conducting stripes. We have recently\cite{nature}
developed a theory of the phase diagram of fluctuating conducting  
stripes for doped Mott insulators.
In the present context, the ``edge states" 
at the interface between two regions of differently quantized Hall 
conductance play the same role as the conducting stripes.  However, 
there are two 
important differences between these edge states and the fluctuating 
stripes we studied in 
the context of doped Mott insulators:  i) because of the high 
magnetic 
field, 
the edge states are intrinsically chiral (with alternating 
chirality, as 
shown in Fig.\ \ref{fig:fig1b}), and ii) the high temperature phase 
has 
full $U(1)$ rotational 
symmetry, as opposed to the discrete rotational (point-group) 
symmetry 
found in doped insulators.  We will show here that the  quantum
mechanical fluctuations about  
the Hartree-Fock state 
of references \ref{bib:kfs} and \ref{bib:chalker} 
lead to a variety of new phases.
Based on these results, we propose the qualitative
$T=0$ phase diagram in the absence of
disorder shown in Fig.\ \ref{fig:fig2}.
 
This phase diagram  includes the following
{\sl electronic liquid
crystalline} and true crystalline phases:
\begin{enumerate}
\item
{\sl A quantum smectic}, which has charge-density wave order 
which breaks translational and rotational symmetry, but in which the 
liquid-like (metallic) behavior of the chiral edge states is 
preserved.  
\item
{\sl A quantum nematic}, in which quantum fluctuations of the stripe 
order 
are sufficiently strong to restore translational symmetry, {\it 
i.e.} to melt the CDW order, but still small enough that local 
orientational order of the stripes persists, thus breaking rotational 
symmetry.  
\item{\sl A quantum isotropic fluid phase}, in which the 2DEG is 
invariant 
under both rotations and translations. 
\item{\sl An insulating stripe-crystal} phase is
also possible, when the partially filled Landau level is not 
half-filled (\ie is not 
particle-hole symmetric). This phase is characterized by the same 
CDW order 
perpendicular to the stripe direction as the smectic, but in addition
the density wave fluctuations along neighboring 
stripes phase-lock to each other, forming a true, insulating, two-dimensional 
electron crystal.
\item{\sl A Wigner Crysal} phase is also expected, especially at 
partial filling of the Landau level near 0 or 1.  This phase is also 
insulating, but differs from the insulating stripe crystal in its 
crystal structure, and its degree of isotropy.
\end{enumerate} 



The smectic and the nematic phases both break rotational 
symmetry;  because of the conducting character of the chiral 
edge states, both liquid crystalline phases
possess highly anisotropic conductivity tensors, with principal 
axes parallel to (``x-direction'') and perpendicular to 
(``y-direction'') the preferred stripe orientational direction.  
Both phases are compressible and have 
a non-quantized Hall conductance. 
The stripe and Wigner crystal phases are ``insulating'', since the
crystals are easily pinned by impurities or boundary effects.
Of course, what this means is that the full conductivity 
tensor at low temperature is that of the lower-lying, full Landau 
levels, so these states are actually quantized Hall states.  The 
isotropic fluid is actually a set of phases, 
including the fractional quantum Hall and 
compressible Hall metal phases which are familiar from previous 
studies. 

Because 
$d=3$ is the lower critical dimension for smectic order,\cite{lcd}
the electron smectic only exists at $T=0$;  at finite temperature, it 
is indistinguishable from the nematic. 
Thermal fluctuations 
also eliminate true long-ranged 
orientational nematic order, but quasi-long-ranged (power-law) 
order 
survive up to a finite temperature transition.  Wigner 
crystalline order 
is destroyed by thermal fluctuations in the standard (2D melting) 
fashion, {\it i.\ e.\/} either there is a single, first-order melting
transition, or a sequence of two continuous transitions and
an intermediate temperature ``hexatic'' phase, 
with short-range  positional and six-fold orientational 
quasi-long-range order.\cite{2dmelt}  The stripe-crystal order is destroyed in a 
similar fashion, although the intermediate state phase in this case 
is a ``biatic'', which from a symmetry point of view is 
indistinguishable 
from the finite temperature nematic. (See Fig. 3.)

The paper is organized as follows. In section \ref{sec:mft} we 
summarize the main results of the mean field theory of Refs.
 \ref{bib:kfs} and \ref{bib:chalker}.
Here we argue that this 
state should be viewed as an electron smectic. 
In section \ref{sec:qf} we discuss in detail the 
effects of quantum fluctuations and 
their implications for the $T=0$ phase
diagram. We focus on two main effects: fluctuations 
of the geometry (or shape) of the stripes and intra-stripe charge 
density fluctuations. We show that
intra-stripe fluctuations alone always produce an
instablity of the smectic to a stripe-crystal
phase.  In contrast, small shape fluctuations tend to stabilize
the smectic.\cite{nature} Furthermore, we argue that large shape 
fluctuations lead to a two stage quantum melting of the smectic 
through an intermediate nematic phase into an isotropic electron 
fluid.
In this section we also characterize the various phases in the phase 
diagram.
In section \ref{sec:FT} we discuss the effects of 
thermal fluctuations and the fate of the $T=0$ phases.
In section IV, we discuss on general grounds the expected behavior of 
the conductivity tensor in the
various phases, especially its behavior and anisotropies
at low temperatures.   Section  V contains a brief and highly 
incomplete discussion of the effects of quenched randomness 
(disorder) on the principal finings of this paper.  Finally, in 
section \ref{sec:exp} we discuss the relation between this 
theoretical picture and the recent experiments of Lilly, Cooper, 
Eisenstein, Pfeiffer and West\cite{jim1}.

\section{Mean-field theory and the smectic phase}
\label{sec:mft}

\begin{figure}
\begin{center}
\leavevmode
\epsfxsize=2in
\epsfbox{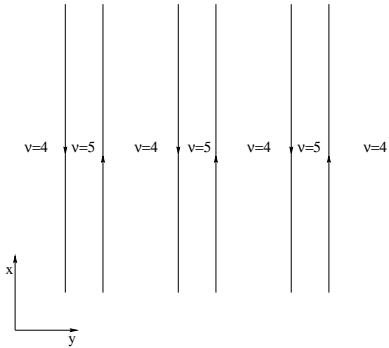}
\end{center}
\caption
{Schematic view of the smectic phase:  In this picture, we have taken 
the filling factor somewhere between $\nu=9/2$ and $\nu=4$.
The system is compressible with the fraction of the sample with 
filling fraction $\nu=5$
decreasing with increasing magnetic field.  As this pogresses,
nearby pairs of edge states become strongly coupled and the result is 
a
stripe (smectic) phase of non-chiral Luttinger liquids separating 
regions with
$\nu=4$. 
This phase exists only at $T=0$.
}
\label{fig:fig1b}
\end{figure}

We take as our starting point a mean-field state
which consists of alternating stripes with filling 
fraction
$\nu=M$ and $\nu=M+1$
as shown in Fig.\ \ref{fig:fig1b} for the case $M=4$.
Here, we have assumed, for simplicity, that the cyclotron and Zeeman 
energies are 
sufficiently large that electrons sequentially fill individual 
spin-polarized 
Landau  levels as $B$ decreases, and
$M=2N$ or $2N+1$ depending on whether 
the partially filled spin-polarized level is spin up or spin down.  
The striped-CDW state is 
charcterized by an order parameter $\Delta_{CDW}$ which describes a 
charge density modulation with wavelength $\lambda$. In the 
Hartree-Fock description of Refs. \ref{bib:kfs} and \ref{bib:chalker},
the single-particle states in 
the $N$-th Landau level near the ``crest" of the CDW are 
filled while those near the ``trough" are empty. 
The wavelength $\lambda$  of the CDW 
is of the order of
$R_c$, the cyclotron radius  in the $N$-th Landau level, $\lambda 
= A R_c= A \ell_0
{\sqrt{N}}$, where $\ell_0={\sqrt{{\frac{\hbar c}{e B}}}}$ is the 
magnetic length. Here
$A$ is a constant determined by the details of the interaction and 
$A$ 
increases
smoothly as the interactions become progressively screened. For $N 
\gg 1$, the Fermi
wavelength $\lambda_F$ is small, $\lambda_F \ll R_c$. 


It follows from general hydrodynamic principles that there exist 
gapless edge states at the boundary between two regions of 
differently 
quantized Hall conductance.\cite{halperin,wen,stone}  In the case of 
a boundary between two integer quantized Hall states, these 
hydrodynamic 
edge modes can be simply constructed as particle-hole excitations, 
which propagate with a velocity, $v= c E_{edge}/B$, which is 
proportional to the strength of the electric field at the edge, 
$E_{edge}$.
Thus, in the absence of interactions between edges, these excitations
form {\sl chiral Fermi liquids}, and intra-edge
electron-electron interactions only renormalize 
the velocity $v$.  
In an ordered 
stripe phase, there are two such chiral edge states per unit cell 
with 
oposite chirality, as shown in Fig.\ \ref{fig:fig1b}.  

However, there is an important distinction between
the edge states that occur on the boundaries of quantum 
Hall devices, which lie along equal-potential contours
defined by an externally applied gate voltage, and the 
internal edge states in a stripe phase where the edges are
self-consistently generated. In the latter case,
in addition to the intra-edge excitations described above, 
there is a second class of low 
energy excitations associated with deformations of the 
effective potential itself, or in other words with the ``shape" (and 
even topology) of the stripe structure. Formally, 
the Hartree-Fock state can be thought of as a saddle-point 
solution of an imaginary time path-integral in which an effective 
potential has been introduced as a Hubbard-Stratonovich field, 
$\Delta_{CDW}(\vec r,t)$, which is just the local CDW order 
parameter.  
The intra-edge 
excitations occur with fixed $\Delta_{CDW}(\vec r,t)$,   
while the shape excitations involve defomations 
of $\Delta_{CDW}(\vec r,t)$, itself. A 
uniform order parameter $\Delta_{CDW}$ defines an ordered CDW state 
with
wavelength $\lambda$. Because this is a state of spontaneously 
broken symmetry, the transverse excitations are Goldstone modes, and hence gapless.
These are the stripe deformations referred to abobe.
(It is sometimes useful to think of the intra-edge 
excitations as the ``quasi-Goldstone modes'' associated 
with an almost broken translational symmetry, \ie the quasi-long range 
order along the stripe direction.\cite{multi})

When $\nu=M+1/2$, the system is particle-hole symmetric, which
means that exactly half of the area is occupied by regions of $\nu=M$
and half by regions of 
$\nu=M+1$ integer quantum
Hall liquid.  As the magnetic field is increased so that
$\nu$ varies between $M+1/2\ge \nu \ge M$, 
the ratio of areas of the two locally 
coexisting 
quantum Hall liquids varies from 0 to 1.  (The range $M+1\ge \nu \ge 
M+1/2$
is related to the range $M+1/2\ge \nu \ge M$ by particle-hole 
symmetry.)
If we denote by $D_{M}$
the width of each strip of $\nu=M$,
then the ratio of $D_{M}/D_{M+1}$ is determined by the
filling fraction according to
\be
\nu-(M+{\frac{1}{2}})
=\frac{D_{M}-D_{M+1}}{D_{M}+D_{M+1}}.  
\ee
The length scale $\lambda = {D_{M}+D_{M+1}}$ is 
determined by 
the competition between the short and long-range pieces of the 
Coulomb potential, and so depends on the details of the 
short-distance screening - in the schematic phase diagram in Fig.\ 
\ref{fig:fig2}, 
the y-axis signifies appropriate changes in the short-range piece of 
the Coulomb interaction, which on a phenomenological level we roughly 
associate with changes in Landau-level index, $N$. 

In the next section, we will consider the effects of interactions 
between the edge states.  As $\nu$ is decreased from $M+1/2$, 
the edge states (of opposite
chiralities) on either side of each $\nu=M+1$  
stripe begin to approach each other;  $D_{M+1}$ decreases.
Consequently, the
interactions between  these pairs of edges grow stronger 
and the interactions among the electrons in different pairs of edges 
(separated by $D_{M}$)
grow weaker. For filling 
fractions $\nu \neq M+{\frac{1}{2}}$,  rather then thinking of 
individual, chiral edges, we should consider the excitations of an 
array of non-chiral one-dimensional structures, 
each consistituted from a {\it pair} of chiral edge states.

At zero temperature, mean-field 
stripe ordered states have a stripe spacing which varies continuously 
with $\nu$, so they are clearly compressible.  
They spontaneously break the $U(1)$
rotational invariance of the 2DEG as well as translation invariance 
along one direction.
Thus, these states have both orientational and translational long 
range order (in one
direction); this is an {\sl electron 
smectic}.  As a consequence, we expect $\sigma_{xy}\sim nec/B$ to be 
unquantized, and $\sigma_{xx}> \sigma_{yy}$;  indeed,
we show below that, in 
the absence of disorder, $\sigma_{xy}= \sigma_{yx}=e^{2}\nu/h$, 
$\sigma_{xx}$ diverges, and $\sigma_{yy}$
vanishes as $T\rightarrow 0$.
In addition, at precisely $\nu=M+{\frac{1}{2}}$, where
the system has an exact particle-hole symmetry (in
the half-filled spin-polarized Landau level), this discrete ${\bf 
Z}_2$ symmetry 
is also spontaneously broken.

\section{Effects of Quantum Fluctuations}
\label{sec:qf}

So far we have ignored the effects of quantum and thermal
fluctuations around the mean-field state. Two distinct sorts of 
quantum fluctuation 
effects can fundamentlly change the character of the  
ground-state: 1) fluctuations of the interacting one-dimensional 
metallic intra-edge degrees of freedom (induced by electron-electron 
interactions), and
2) shape fluctuations 
in the positions, and ultimately even the connectivity, of the edges 
themselves.\cite{nature}

The fluctuations of the metallic edge degrees of freedom can be 
described most simply
using standard bosonization methods.  At $\nu=M+{\frac{1}{2}}$, 
the low energy charged degrees of
freeedom that are active in the smectic phase are the fluctuations of 
the ``edge states", 
which are
described by an array (with alternating chiralities) of Fermi 
liquids, 
\ie  chiral bosons with
unit compactification radius (or Luttinger parameter). Well away 
from  
$\nu=M+{\frac{1}{2}}$, the
charged degrees of freedom of each {\it pair} of close-by edges
form a {\sl non-chiral Luttinger liquid} with a Luttinger parameter 
$K<1$, 
which is a smooth function of the strength of the intra-pair Coulomb 
repulsion, which 
is in turn a function
of the mean separation between the edges in a given pair, \ie of 
$D_{M}$ for $\nu< M+1/2$.   
Note\cite{nature} that for $K<1$, the zero temperature density fluctuations 
associated with an isolated pair of edges exhibit quasi-long-range order,
\be
<{\cal O}_{j}^{\dagger}(x){\cal O}_{j}(0)> \sim \cos(2k_{F}^{eff}x + 
\theta_{0})/|x|^{2K}
\ee
where ${\cal O}_{j}$ is the $2k_{F}^{eff}$ piece of the 
charge-density\cite{emery} 
operator on the $j^{th}$ stripe ({\it i.e.} the  $j^{th}$ {\it pair} of 
edge states).
Consequently the intra-pair
CDW susceptibility diverges as the temperature 
$T\rightarrow 0$ as 
\be
\chi_{CDW}\sim T^{-2(1-K)}
\ee
where the CDW has a period 
determined by an effective 
value of 
\be
2k_{F}^{eff}=D_{M}/\ell_0^{2}.
\ee

Direct electron tunneling, and even pair tunnelling between pairs of ideal 
straight edges are forbidden by momentum 
conservation. However, for $\nu$ not too close to $M+1/2$, the 
Coulomb interactions between 
neighboring pairs of edges couples the intra-pair CDW fluctuations;  
schematically, this makes a
contribution to the Hamiltonian density 
\be
{\cal H}_{Coul}= V \left\{ {\cal O}_{j}^{\dagger}(x){\cal 
O}_{j+1}(x) 
+ {\rm H.c.}\right\}.
\ee
The scaling dimension of the operator ${\cal H}_{Coul}$ which represents 
the coupling of the $2k_{F}^{eff}$ CDW order 
parameters on neighboring stripes is $d_{CDW}=2K$. For repulsive 
Coulomb interactions $K<1$, 
$\chi_{CDW}$ diverges and $d_{CDW}< 2$. In the 
renormalization group sense, this coupling 
is relevant. It produces an
instability of the smectic phase, analagous to one
that is commonly encountered in 
{\it quasi}-one dimensional materials, toward the formation of an 
insulating, stripe crystal phase with long-range CDW 
order both along and transverse to the stripe direction.

This is not the whole story, since, as discussed above, the stripes 
are 
spontaneously generated,
so their {\sl shapes} 
are also dynamically fluctuating low energy degrees of 
freedom\cite{nature}. 
Moreover, the
couplings between these geometric degrees of freedom and the 
(bosonized) charge
fluctuations are non-linear and involve many 
derivatives\cite{comment}. It is somewhat technical but nevertheless
possible to show
\cite{paper} that backscattering processes assisted by weak shape 
fluctuations of the
stripes leads only to further renormalizations of the Luttinger 
parameter.  
Where even weak shape fluctuations can be qualitatively 
important 
is through their effect on the CDW fluctuations on neighboring pairs 
of edges. Since the fluctuating CDW order oscillates with an 
effective Fermi
wavelength $\lambda_F=2\pi/2k_{F}^{eff}$ along the {\it locally 
defined }
stripe direction, the slightly different geometries defined by 
neighboring stripes means that the CDW fluctutuations on those 
stripes 
are geometrically dephased when the arc-lengths differ by an amount 
of order $\lambda_F$. Formally, these fluctuations induce an 
additional phase factor, 
\begin{eqnarray}
{\cal O}_{j}^{\dagger}(x){\cal 
O}_{j+1}(x) && \rightarrow {\cal O}_{j}^{\dagger}(x){\cal 
O}_{j+1}(x)  \exp\{i2k_{F}^{eff} \Delta_jL(x) \}
\nonumber \\
&&
\label{eq:coupling}
\end{eqnarray}
in the expression for ${\cal H}_{Coul}$ where $L_{j}(x)$ is the arc 
length to position $x$ measured along the $j^{th}$ stripe.
Here we have defined $\Delta_jL(x)=L_{j}(x)-L_{j+1}(x)$.
In reference [\ref{bib:nature}] we showed that this sort of 
fluctuation renders the coupling between CDW fluctuations on 
neighboring stripes irrelevant, and this produces a first-order 
transition as a function of the magnitude of the shape 
fluctuations from an insulating stripe-crystal phase for small 
fluctuations to a conducting smectic phase for larger ones.  
Naturally, the smaller $\lambda$, or 
equivalently the closer $\nu$ is to $M+1/2$, the more sensitive the 
CDW
ordering is to small amplitude shape fluctuations. As a 
consequence, we generally expect the smectic phase to be more stable 
for $\nu$ near $M+1/2$ and the stripe crystal phase to be more stable 
away from this value.

So far, we have only considered the case in which the shape 
fluctuations are sufficiently small that they do not damage the basic 
stripe order of the mean-field ground-state.  
When the shape fluctuations  grow in magnitude to be comparable to the 
spacing 
between edges, backscattering interactions
assisted by non-linear retarded shape fluctuations induce operators 
that break up the
stripes.
In other words, these operators generate {\sl dislocations} in the 
smectic stripe
order. These operators are irrelevant at weak coupling. The strength 
of this coupling is a measure of the effects
of quantum fluctuations on the stripe structure and
it decreases with increasing stripe
rigidity\cite{nature}.
When these operators 
become relevant, the
system undergoes a {\sl quantum phase transition} from the {\sl 
smectic} state 
to the {\sl nematic} state in which dislocations proliferate. In this 
state
rotational long range order is still present but translation 
invariance is restored.
Since this phase is far from the mean-field state from which we 
started, our knowledge of its properties is less certain.  However, 
because in two spatial dimensions, the smectic to nematic phase 
transition is 
expected, from Landau theory, to be continuous, we can imagine that 
substantial local stripe order persists well into the nematic phase.  
As a consequence we expect this phase to be compressible, to posses a 
non-quantized Hall conductance, and an anisotropic longitudinal 
conductivity with $\sigma_{xx}> \sigma_{yy}$.  

A schematic $T=0$ phase diagram which summarizes the above 
considerations 
is shown in Fig.\ \ref{fig:fig2}.  Here, the x-axis is the partial 
filling of the 
highest occupied spin-polarized
Landau level and the y-axis is a microscopic quantum parameter, 
related 
to the strength of the short-range piece of the Coulomb interaction, 
which determines the typical magnitude of shape 
fluctuations in units of the stripe width.  (Roughly, since 
fluctuations about the mean-field state are thought to become less 
severe with increasing Landau index, $N$, we have identified this 
coordinate with $1/N$, with shape fluctuations of the stripes being 
increasingly important the larger $1/N$.)   

The general structure of the phase 
diagram along its edges is completely determined by general 
principles and the above 
considerations.  In the vicinity of the $\nu=M$ axis, the system can be
thought of as consisting of dilute quasi-particles, which thus 
necessarily form a triangular lattice quasi-particle Wigner crystal\cite{foot}.  
This is separated by a line of first-order transitions from the 
various phases discussed in this paper.  The instability of 
the smectic phase to stripe-crystal order in the absence of shape 
fluctuations means that along the x-axis, the system is always 
crystalline;  a stripe crystal for $\nu$ near $M+1/2$ and a Wigner 
crystal for smaller $\nu$.  For $\nu=M+1/2$, the smectic phase is 
marginally stable, due to particle-hole symmetry, or equivalently, 
due to the fact that the edge states here are chiral Fermi liquids.  
However, as the quantum fluctuations become more severe, the smectic 
phase melts, first to form a stripe nematic and then an isotropic 
liquid phase.  Presumably, for $\nu=M+1/2$,
this isotropic state is the famous Hall metal.  
Finally, for large quantum parameter and variable $\nu$, the system 
is dominated by the 
familiar liquid states, including various Hall metal and fractional 
quantum Hall 
liquid states;  we group all these states into one region of the 
phase 
diagram labelled ``isotropic''.  Within these constraints, we have  
used artistic license to 
complete the schematic phase diagram in a consistent manner.  The 
resulting phase diagram bears strong similarities with the 
phase 
diagram we constructed previously for doped 
antiferromagnets.\cite{nature}

Associated with the various broken continuous symmetries of the $T=0$
phases are a set of Goldstone modes whose character can be deduced 
from general principles.  Detecting these modes may, ultimately, be 
the most direct way of unambiguously identifying the various broken 
symmetry phases experimentally.  

The existence of Goldstone modes follows directly from the 
generalized 
elastic theory of the electron liquid crystal phases.
However, since the magnetic field affects the 
dynamics of charge motion, the modes
studied here have quite different character than those of the 
corresponding phases in zero magnetic field.  Moreover, because of the
high density of low energy charge excitations associated with the 
edge states, 
dissipation may play a significant role 
in the dynamics of the Goldstone modes. 

We expect the Goldstone modes of the crystalline phases to be fairly 
standard. In particular, the Wigner crystal phase has been 
extensively studied in the literature\cite{wc}. The stripe crystal 
phase should be analogous except for differences due to the different point 
group symmetries of the two crystalline structures.

The liquid crystal phases have not been studied so far. The
transverse fluctuations of the CDW structure are the Goldstone 
modes. These modes per se do not couple directly to charge 
fluctuations. However, the locus of the nodes of the CDW  
modulate the local charge density profile and through it they 
determine the local structure of the edge states. These two sets of 
low energy modes, local fluctuations of the stripes and charge 
fluctuations along each stripe, govern the low energy physics of the 
smectic phase. In a separate publication we will present a theory of 
these modes. 

\section{Finite Temperature Effects}
\label{sec:FT}

At finite temperature  continuous symmetries cannot be broken
in two dimensions, so both translation symmetry (which is 
spontaneously broken at $T=0$ in the smectic 
and the two crystalline phases) and rotational symmetry  (in 
all liquid crystalline and true 
crystalline phases) are clearly restored for $T>0$.
However, quasi-long-range order is possible, even at 
finite temperature, so phases can still be distinguished according to
what power-law order they possess.  

The smectic phase is destroyed at finite 
temperature due to a
proliferation of dislocations;  three is the lower critical dimension 
for smectic order\cite{lcd}.  This is even true in the presence of 
long-range 
Coulomb interactions.\cite{zohar}

A nematic phase, with power-law orientational 
order, survives to finite temperature.  Indeed, we 
expect that everywhere in the $T=0$ phase diagram where smectic or
nematic order exists, quasi-long ranged orientational order will 
survive
for small $T>0$.
In addition, by analogy with the 
theory of
hexatic phases in two dimensions,\cite{2dmelt} there should be a finite 
temperature phase
transition to a fully isotropic 2DEG mediated by unbinding of 
disclinations.  Above this temperature, all tensor quantities, such as
the conductivity, should be isotropic, while below it, anisotropies 
will develop which will grow with decreasing temperature. 


\begin{figure}
\begin{center}
\leavevmode
\epsfxsize=3in
\epsfbox{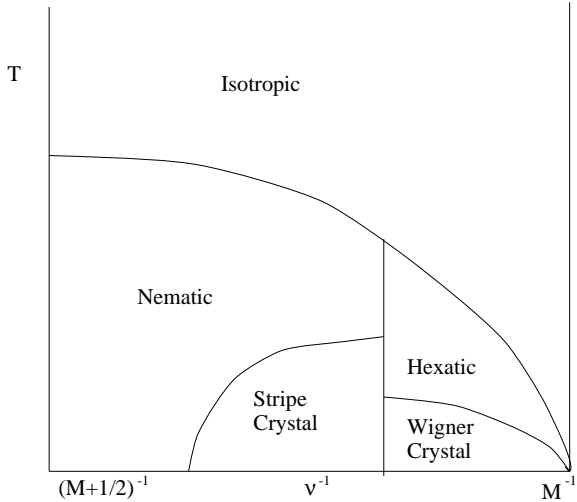}
\end{center}
\caption
{Schematic finite temperature phase diagram as a function of inverse 
filling factor along line c of figure 1.}
\label{fig:fig4}
\end{figure}


Both crystalline phases will melt at finite $T$ by one of the more or 
less standard routes for two dimensional melting, that is either via 
a first order transition, or by a sequence of two transitions. In the 
latter case, there will be a low temperature solid phase with 
power-law
positional and orientational order, and a 
non-vanishing shear modulus.  This solid phase melts via a dislocation
unbinding transition to an intermediate (liquid crystalline) state 
with power-law orientational 
and short-range positional order.  In the case of the melting of the
stripe crystal, this ``biatic'' phase is not fundamentally
distinct from the 
finite temperature nematic phase discussed above, although in 
practice, the melted crystal may still be fairly insulating, 
whereas the 
nematic is moderately conducting.  In the case of the Wigner 
crystal, the intermediate phase is a ``hexatic'', in which the 
power-law orientational order has a six-fold rotational symmetry, 
rather than the two-fold symmetry of the nematic.  As indicated 
above, 
these  intermediate phases 
give way to a fully isotropic high temperature phase via a continuous
disclination unbinding transition.

In all critical phases, which is to say all the finite temperature 
phases described above, the effects of symmetry 
breaking fields are particularly dramatic.  At low temperatures, and 
in the absence of such symmetry breaking fields, there is no true 
broken symmetry and no order-parameter. In these phases the 
system is actually in a critical region terminating at a 
Kosterlitz-Thouless phase transition at a critical temperature $T_c$. 
Naturally, this phase transition is rounded by a symmetry breaking 
field. However, below $T_c$, even a small symmetry 
breaking field, $h$, produces a large response.  This intuition is 
made precise in the sense that the exponent, $\delta >1$, where
\be
m \sim |h|^{1/\delta}
\label{eq:m}
\ee
where $m$ is the value of the order parameter.  For instance, for the 
nematic phase, we can define $m={\sigma_{xx}-\sigma_{yy}\over 
\sigma_{xx}+\sigma_{yy}}$, and $h$ 
then 
is a dimensionless measure of the underlying anisotropy of the 
substrate.  
Near the 
critical temperature, $\delta$ approaches the universal value 15, 
and $\delta$ diverges as $T\rightarrow 0$.  As a function of temperature, 
this exponent can be computed exactly in terms of 
the anomalous dimension, $\eta$, of the XY model
(or, with the same result, from a self-consistent
phonon approximation): 
\be
\delta= {\frac{4}{\eta}}- 1=\frac {2 \pi \kappa(T)} T -1
\label{eq:delta}
\ee
where $\eta$ approaches 1/4
as $T\rightarrow T_{c}$.  Here, 
$\kappa(T)$ is the long wave-length helicity modulus,
which approaches a constant as $T\rightarrow 0$, and the universal 
value $\kappa(T_{c})=4T_{c}/\pi$
as $T\rightarrow T_{c}$. (There is 
a factor of $4=2^{2}$ difference here than in the usual XY model since the 
vortices in a nematic have half the usual topological charge.) As $T \to T_c$,  
$\delta$ reaches the universal value $\delta=15$.
 The large values of $\delta$ imply that very small microscopic anisotropies 
have an enormous orienting effect. 

\section{The Conductivity Tensor}
\label{sec:cond}

In this section, we discuss the expected behavior of the conductivity 
tensor in the various parts of the phase diagram in Fig. 1.  
It follows directly from the Kubo formula, 
so long as there are no singularities (such as are found in a 
superconductor) associated with the $\vec k$ and $\omega \rightarrow 
0$ limit, that $\sigma_{xy}=\sigma_{yx}$ (and, consequently, that 
$\rho_{xy}=\rho_{yx}$).  This statement is true independent of 
whether or not the system is rotationally symmetric.  In 
general, in a state with a four-fold rotational symmetry (which, of 
course, includes all isotropic liquid states) 
$\sigma_{xx}=\sigma_{yy}$.  Conversely, in any state 
which is not four-fold rotationally symmetric and has a finite Hall 
conductance, there is no {\it a priori} reason to expect this 
equality to 
hold, and therefore there is every reason to expect it not to hold.  
Thus, in all the crystalline and liquid crystalline states discussed 
here, we expect $\sigma_{xx}\ne \sigma_{yy}$, although for the two 
crystalline states, since they are insulating, we expect that both 
diagonal components of the conductivity tensor will be small at low 
tempertures.

\subsection{The Smectic Phase}

At $T=0$ and
in the absence of impurities, the smectic phase is boost invariant 
in the stripe direction.  As a consequence, under conditions in which 
the electric field is perpendicular to the stripe direction, it is 
possible to go to a co-moving frame in which the electric field 
vanishes. It therefore follows that 
the Hall conductance tracks the change in
the filling fraction, $\sigma_{xy}=\nu {\frac{e^2}{h}}$, and that
$\sigma_{yy}=0$.  
On the other hand, because there are a finite density of conducting 
channels, and because of the irrelevance of all 
back-scattering interactions in the smectic phase, the longitudinal 
conductivity in the stripe direction, $\sigma_{xx}$, 
diverges in the limit $T\rightarrow 0$.  Impurities will, of course, 
alter these conclusions, but for weak disorder and low but non-zero 
temperature, one would still expect $\sigma_{xy}\sim e^{2}\nu/h$ and
$\sigma_{xx} \gg e^{2}/h \gg \sigma_{yy}$.

~From a microscopic viewpoint, if an
external electric field {\sl perpendicular} to the stripes is 
applied, every stripe with
filling fraction $\nu_{M}$ ($\nu_{M+1}$) has an induced Hall current 
{\sl parallel} to the
direction of the stripe and the Hall conductance of that stripe is 
$\sigma_{xy}(M)$
($\sigma_{xy}(M+1)$). Notice that in this configuration the current 
of the edge states
separating each pair of nearby stripes is part of the Hall current. 
The Hall current
changes continuously and no longitudinal current is induced in this 
configuration. However, if
the external electric field is applied {\sl parallel} to the stripes, 
the induced Hall
current in each stripe is now {\sl perpendicular} to the stripes. 
(Recall that nearest
neighboring stripes have different Hall conductance.) As the boundary 
between two stripes is
approached the current switches from being perpendicular to the 
stripe to being parallel to
the stripe and it is carried by the corresponding edge state. 
Thus, current is conserved
but, in addition to the ``bulk" Hall current (which is the same as in 
the other
configuration) there is now a current {\sl parallel} to the external 
electric field and it is
carried entirely by the edge states. 
At least at the mean-field level, 
$\sigma_{xx}\sim C e^{2}/h$ where $C$ is the number of 
edges that make it across the system (and so diverges with the size 
of the sytem) while $\sigma_{yy}=0$. 
This result is correct for a clean system and it is 
robust against quantum fluctuations provided that 
they do not destablize the smectic phase. 

\subsection{The Nematic Phase}

Because the nematic phase is featureless, and so boost invariant, it 
follows that at $T=0$, $\sigma_{xx}=\sigma_{yy}=0$ and
$\sigma_{xy}=e^{2}\nu/h$.  For $T >0$, since the nematic
is a critical phase, the zero temperature result is likely to 
be strongly modified.  On general dimensional grounds, it is 
reasonable to expect $\sigma_{xx}\sim e^{2}/h \gg \sigma_{yy}$, and 
that $\sigma_{xx}$ is greatest at $\nu=M+1/2$, where the Luttinger 
exponent $K$ associated with the edge states is largest, and drops 
symmetrically (due to particle-hole symmetry) as $\nu$ is varied from 
this value.  There is no reason to expect $\sigma_{xy}$ to be 
strongly temperature dependent.

\section{Effects of Quenched Disorder}
\label{sec:dis}

Disorder likely eliminates most of the sharp distinctions between 
phases, and hence turns most of the phase transitions discussed above 
into crossovers.  However, if the disorder is sufficiently weak, then 
the crossovers can be sharply defined, and important local 
distinctions between the various ``phases'' should be experimentally 
detectable.  Certainly, neither broken translational nor rotational 
symmetry survive disorder.  

The effects of disorder on the 
conductivity tensor in the $T\rightarrow 0$ limit are likely to be 
severe and non-perturbative;  even weak disorder can cause 
localization.  However, at non-vanishing temperatures, we can expect 
that in the low disorder limit, the conductivity tensor will resemble 
that of the ideal system.  Interesting non-linear effects, involving 
pinning of the various forms of CDW order, can be expected in the 
presence of disorder.  In the stripe crystal and Wigner crystal 
cases, 
these effects are well studied previously, but for the smectic they 
may have some novel features.  Because the Luttinger exponent $K < 
1$, disorder is a relevant perturbation to the one-dimensional 
edge state problem in the absence of stripe shape fluctuations.
Thus, disorder is likely to produce
dramatic decreases in the diagonal matrix elements of the 
conductivity tensor at sufficiently low temperatures in both 
electronic liquid cryalline phases.
In general, the effects of weak 
disorder on electronic liquid crystals is an area for future study. 

\section{Relation to Experiments}
\label{sec:exp}

The study undertaken in the present paper was originally
motivated by some very recent and remarkable experiments 
done by  Lilly, 
Cooper, Eisenstein, Pfeiffer and West\cite{jim1} in which large, 
temperature dependent anisotropies were discovered in 
the 2DEG under conditions in which 2 (or more) Landau levels are full.
The experiments were 
done in ultra-high
mobility $GaAs/AlGaAs$ heterojunctions. That the 
samples have very
weak disorder is indicated, for instance, by the observation of the 
quite
fragile quantum Hall plateau at $\nu={\frac{5}{2}}$ and by the large 
number of fractional
quantum Hall states seen in the lowest Landau level. The salient 
features 
of the experiments
of Ref. \ref{bib:jim1} are as follows:
For a partially filled third Landau level with $\nu$ in
the neighborhood of $\nu= {\frac{9}{2}}$, 
there is a characteristic
temperature $T_0 \approx 150 mK$ above which the resistivity tensor is
nearly isotropic, and below which there is  a rapid 
crossover 
to a highly anisotropic 
compressible state. The resistivity tensor in this state exhibits a 
non-quantized Hall resistance and an anisotropic longitudinal response, such 
that
$\rho_{xx}$  
grows very rapidly with decreasing temperature
until it reaches a value of order $1000 \Omega$ at the lowest 
temperatures, $T\approx 25mK$, while $\rho_{yy}$, measured by
rotating the current by $90^\circ$, becomes very small.  
Since in the abscence of disorder, on theoretical grounds 
$\rho_{xy}=\rho_{yx}= 
(h/e^{2})\nu^{-1}$, we expect that this relation should hold
approximately in this system.
Inverting 
this tensor we find, for the conductivity tensor at low temperatures,
$\sigma_{yy}\sim e^{2}/h$, $\sigma_{xy}=\sigma_{yx}\sim \nu e^{2}/h $, 
and $\sigma_{xx}$ small.  The low temperature value of 
$\rho_{xx}$ as a function of $\nu$ exhibits a broad peak centered at 
$\nu=9/2$, with a width in $\nu$ which is 
substantial and 
approximately temperature independent.
Structure is also seen in the 
``wings" of
the Landau level
even at $150 mK$. Specifically, two very well defined pairs of quantum
Hall plateaus are seen, one pair with $\sigma_{xy}=4e^{2}/h$ for 
$\nu$ near 4, and another one with $\sigma_{xy}=5e^{2}/h$ for $\nu$ 
near 5. However, the resistivity peak between the two plateaus (with the
same quantized Hall conductance) becomes smaller with decreasing
temperature, in contrast with the usual critical peaks seen in
transitions between plateaus. 

The same structure is repeated for $\nu$ in the vicinity of 11/2, 
13/2, 15/2, and beyond, although it apparently becomes more difficult 
to resolve beyond $\nu=15/2$. In particular, the peak value of 
$\rho_{xx}$ 
decreases with increasing $\nu$, roughly in such a way that 
$\sigma_{yy}$ at $\nu=(2M+1)/2$ remains in the vicinity of $e^{2}/h$.
It is important to stress that, even at low temperatures, no 
substantial anisotropy is apparent at the lowest temperatures
at smaller values of $\nu$, in 
particular near $\nu=7/2$ and $5/2$, nor at any magnetic field at 
temperatures in excess of 100mK.

The existence of this anisotropy observed in highly pure
samples clearly indicates that this effect 
is driven by
electron-electron interactions and that disorder plays a 
secondary role. This is doubly remarkable since the natural 
expectaion was that precisely in the middle of the plateau there 
should be a phase transition from a $\nu=4$ to a $\nu=5$
quantum Hall liquid. However at the 
transition between plateaus, which is a quantum phase transition 
driven by disorder, although one expects a  peak in $\rho_{xx}$, the 
peak should narrow as $T \to 0$ following a universal scaling law. 
The results of these experiments suggest that, although the system is 
indeed compressible, there is no narrowing of the peak as $T \to 0$. 
Thus the system is critical for a range of filling factors. 
Furthermore the transition between plateaus should be essentially 
isotropic.

The picture presented in this paper gives a natural 
interpretation\cite{jim2} of 
these effects. It is natural to identify the experimentally observed 
anisotropic state with the finite temperature nematic state discussed
above, which is a critical state, and the crossover observed at 
$T\approx 100mK$ with the 
disclination unbinding transition, perhaps somewhat rounded due to 
quenched disorder. A small anisotropy in the 
heterojunction device is the symmetry breaking field which picks a 
preferred orientation for the nematic, and insures 
that there is actual long-ranged orientational order, as described in 
Eq. (\ref{eq:m}).  
The two quantized Hall states observed in the wings 
are naturally identified with the stripe crystal and quasi-particle 
Wigner crystal phases\cite{foot2} that appear as $\nu \rightarrow M$ in the 
phase diagrams in Figs.\ \ref{fig:fig2} and \ref{fig:fig4}.  
It would 
be interesting to test this 
hypothesis by looking for evidence of a finite temperature melting 
transition, or characteristic non-linear I-V's in these ranges of $B$.

This still does not address the question of whether the ground-state 
phase near $\nu=9/2$ is a smectic or nematic.  However, since
the lingitudinal resistivity $\rho_{xx}\sim 1000\Omega$ corresponds to
a longitudinal conductivity 
$\sigma_{yy} \sim 
{\frac{e^2}{h}}$,
it seems likely that the ground-state is either a 
quantum nematic phase or that the stripe order of the underlying 
smectic is strongly disrupted due to pinning by
impurities.
If disorder does not play a dominant role,
the characteristic temperature dependence of the 
nematic order implied by Eq.\ (\ref{eq:delta}) will govern the
resistivity ratio, $\rho_{xx}/\rho_{yy}$.

It is also worth noting that
in the experiments of Willett \etal \cite{willett} in which an 
external modulation was
imposed on a 2DEG in the compressible state at $\nu={\frac{1}{2}}$, 
similar phenomena were observed as in the experiments of Lilly {\it 
et al}.  On the one hand, this gives us greater confidence in 
concluding that the observed anisotropies are a consequence of stripe 
formation.  On the otherhand, it supports the intuitive 
notion\cite{dh}
that the isotropic compressible state at $\nu=1/2$ still has substantial
local stripe correlations, which are simply disordered by quantum 
fluctuations at long distances - this would rationalize the large 
susceptbility of this state to the formation of stripes.  Similarly, 
it may be that anomalously broad 
regions of compressible smectic or nematic phases may 
be stabilized by an externally potential at the ``edge'' of a quantum 
Hall device, producing a form of macroscopic 
edge-reconstruction.\cite{spivak}

\section{ Acknowledgements}

We would like to thank J.\ Eisenstein for bringing this problem to our
attention.
We aknowledge useful discussions with V.~J.~Emery, J.\ Eisenstein,
M.\ Lilly, J.\ Cooper, J.\ Chalker and H-W.\ Jiang.
This work was supported in part by the NSF, grant numbers DMR93-12606 
at UCLA,
NSF DMR94-24511 at UIUC, and NSF PHY94-07194 at ITP-UCSB. 
One of us (EF) is a participant at the ITP Program on
{\it Disorder and Interactions in Quantum Hall and Mesoscopic 
Systems}. EF is grateful to D.\ Gross, Director of Institute for 
Theoretical Physics of the University of California Santa Barbara, 
for his kind hospitality.


\begin{references}
\bibitem{multi}  As is well known, the gapless modes of 
one-dimensional quantum systems with ``spontaneously broken 
continuous symmetries" are not strictly Goldstone modes since these 
symmetries cannot be broken, by virtue of the Mermin-Wagner theorem. 
These systems are actually quantum critical and the gaplessness of 
the modes is a actually consequence of scale invariance.
\bibitem{comment} Strictly speaking these couplings are also 
non-local in time.
\bibitem{foot} Bubble phases which have the symmetry, but different
periodicities, are also possible. See reference [\ref{bib:kfs}].
\bibitem{foot2} In the pure system the transition between the two
crystal states is first order and it is rounded by disorder.
\end{references}
\end{document}